\documentclass[%
twocolumn,
 amsmath,amssymb,
 aps,
pra,
floatfix,
]{revtex4-1} %mohcine   

\usepackage{graphicx}% Include figure files
\usepackage{dcolumn}% Align table columns on decimal point
\usepackage{bm}% bold math
\usepackage[tight,footnotesize]{subfigure}
\usepackage{color,ulem}

\begin{document}

\preprint{APS/GCFM}
\title{Generalized Centrifugal Force Model for Pedestrian Dynamics}% Force line breaks with \\

\author{Mohcine Chraibi}
\author{Armin Seyfried}
 \affiliation{J\"ulich Supercomputing Centre, Forschungszentrum J\"ulich, 
52425 J\"ulich, Germany.}
 \email{{m.chraibi,a.seyfried}@fz-juelich.de}

\author{Andreas Schadschneider}
\affiliation{
 Institute for Theoretical Physics, Universit\"at zu K\"oln, D-50937
K\"oln, Germany
 }%
 \email{as@thp.uni-koeln.de}
\date{\today}% It is always \today, today,

\begin{abstract}
A spatially continuous force-based model for simulating pedestrian
  dynamics is introduced which includes an elliptical volume exclusion
  of pedestrians.  We discuss the phenomena of oscillations and
  overlapping which occur for certain choices of the forces. The main
  intention of this work is the quantitative description of pedestrian
  movement in several geometries. Measurements of the fundamental
  diagram in narrow and wide corridors are performed.  The results of
  the proposed model show good agreement with empirical data obtained
  in controlled experiments.
\end{abstract}

\maketitle

%\tableofcontents

%%%%%%%%%%%%%%%%%%%%%%%%%%%%%%%%%%%%%%%%%%%%%%%%%%%%%%%%%%%%%%%%%%%%%%%%%%

\section{Introduction}

For a beneficial application of pedestrians dynamics, robust and
quantitatively verified models are required.  A wide spectrum of
models has been designed to simulate pedestrian dynamics. Generally
these models can be classified into macroscopic and microscopic
models. In macroscopic models the system is described by mean values
of characteristics of pedestrian streams e.g., density and velocity,
whereas microscopic models consider the movement of individual persons
separately. Microscopic models can be subdivided into several classes
e.g., rule-based and force-based models.  For a detailed discussion,
we refer to~\cite{Schadschneider2009a,Schadschneider2010}. In this
work we focus on spatially continuous force-based models.

Force-based models take Newton's second law of dynamics as a guiding
principle.  Given a pedestrian $i$ with coordinates
$\overrightarrow{R_i}$ we define the set of all pedestrians that
influence pedestrian $i$ at a certain moment as
\begin{equation}
  \mathcal{N}_i:= \{ j :\, \parallel \overrightarrow{R_j} -
  \overrightarrow{R_i} \parallel \le r_c \, \land\, i\,
  \text{``feels''}\, j \}
\end{equation} 
where  $r_c$ is a cutoff radius.
We say pedestrian $i$ ``feels'' pedestrian $j$ if the line joining
their centers of mass does not intersect any obstacle. In a similar
way we define the set of walls or borders that act on pedestrian $i$ as
\begin{equation} 
  \mathcal{W}_i:= \{ w :\, \parallel \overrightarrow{R_{w_i}} -
  \overrightarrow{R_i} \parallel \le r_c \} 
\end{equation}
where $w_i \in w$ is the nearest point on the wall $w$ to the pedestrian $i$.

Thus, the movement of each pedestrian is defined by the equation
of motion  
\begin{equation}
 m_i\ddot{\overrightarrow{R_i}} =
  \overrightarrow{F_{i}} = \overrightarrow{F_{i}^{\rm drv}} +
  \sum_{j\in \mathcal{N}_i} \overrightarrow{F_{ij}^{\rm rep}} +
  \sum_{w\in \mathcal{W}_i} \overrightarrow{F_{iw}^{\rm rep}} \,,
  \label{eq:maineq}
\end{equation}
where $\overrightarrow{F_{ij}^{\rm rep}}$ denotes the repulsive force
from pedestrian $j$ acting on pedestrian $i$,
$\overrightarrow{F_{iw}^{\rm rep}}$ is the repulsive force emerging
from the obstacle $w$ and $\overrightarrow{F_{i}^{\rm drv}}$ is a
driving force.  $m_i$ is the mass of pedestrian $i$.

The repulsive forces model the collision-avoidance performed by
pedestrians and should guarantee a certain volume exclusion for each 
pedestrian. The driving force, on the other hand, models the intention
of a pedestrian to move to some destination and
walk with a certain desired speed.  The set of
equations~(\ref{eq:maineq}) for all pedestrians results in a
high-dimensional system of second order ordinary differential
equations. The time evolution of the positions and velocities of all
pedestrians is obtained by numerical integration.

Most force-based models describe the movement of pedestrians
qualitatively well. Collective phenomena like lane
formation~\cite{Helbing1995,Helbing2004,Yu2005}, oscillations at
bottlenecks~\cite{Helbing1995,Helbing2004}, the ``faster-is-slower''
effect \cite{Lakoba2005,Parisi2007}, clogging at exit
doors~\cite{Helbing2004,Yu2005} are reproduced. 
These achievements indicate that these models are promising candidates
for realistic simulations.  However, a qualitative description is not
sufficient if reliable statements about critical processes, e.g.,
emergency egress, are required. Moreover, implementations of models 
often require additional elements to guarantee
realistic behavior, especially in high density situations.
Here strong overlapping of pedestrians \cite{Lakoba2005,Yu2005}
or negative and high velocities \cite{Helbing1995,Lohner2010} occur which
then has to be rectified by replacing the equation of motion (\ref{eq:maineq})
by other procedures.

Force-based models contain free parameters that can be adequately
calibrated to achieve a good quantitative description 
\cite{HoogendoornDL05,Johansson2007,Kretz2008,Parisi2009,HoogendoornD09}.
However, depending on the simulated geometry the set of parameters
often changes.  In most works quantitative investigations of
pedestrian dynamics were restricted to a specific
scenario or geometry, like one-dimensional motion \cite{Seyfried2006},
behavior at bottlenecks \cite{Kretz2008,Hoogendoorn2003a,Hoogendoorn2005},
two-dimensional motion \cite{Parisi2009} or outflow from
a room \cite{Kirchner2002,KirchnerNS2003,Kirchner2003,Yanagisawa2009}.

In this work we restrict ourselves to corridors and address the possibility 
of describing the movement of pedestrians
in wide and narrow corridors reasonably and in a quantitative manner 
with a unique set of parameters. At the same time, the modelling approach 
should be as simple as possible.

In the next section, we propose such a model which is solely based on the
equation of motion~(\ref{eq:maineq}). Furthermore the model
incorporates free parameters which allow calibration to fit
quantitative data.

%%%%%%%%%%%%%%%%%%%%%%%%%%%%%%%%%%%%%%%%%%%%%%%%%%%%%%%%%%%%%%%%%%%%%%%%%%

\section{The Centrifugal Force Model}
\label{sec:CFM}

The Centrifugal Force Model (CFM)~\cite{Yu2005} takes into account the
distance between pedestrians as well as their relative velocities.
Pedestrians are modelled as circular disks with constant radius. Their
movement is a direct result of superposition of repulsive and driving
forces acting on the center of each pedestrian.  Repulsive forces
acting on pedestrian $i$ from other pedestrians in their neighborhood
and eventually from e.g. walls and stairs to prevent collisions and
overlapping.  The driving force, however, adds a positive term to the
resulting force, to enable movement of pedestrian $i$ in a certain
direction with a given desired speed~$v_{i}^0$.  The mathematical
expression for the driving force is given by
\begin{equation}
\overrightarrow{F_{i}^{\rm drv}} = m_i \frac{\overrightarrow{v_{i}^0} -
 \overrightarrow{v_{i}}} {{\tau}},
\label{eq:fdrv}
\end{equation}
with a time constant ${\tau}$.

Given the direction connecting the positions of pedestrians $i$ and $j$:
\begin{equation}
   \overrightarrow{R_{ij}} = \overrightarrow{R_j}
   -\overrightarrow{R_i} ,\;\;\;\; \qquad
   \overrightarrow{e_{ij}} = \frac{\overrightarrow{R_{ij}}}{R_{ij}}\,
\label{eq:R}
\end{equation}

%%%%%%%%%%%%%%%%%%%%%%%%%%%%%
\begin{figure}
\begin{center}
\includegraphics[width=0.6\columnwidth]{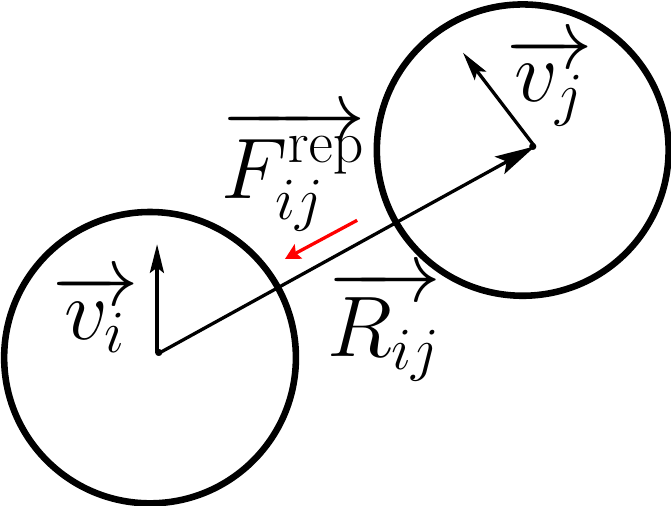}
\caption{(Color online) Direction of the repulsive force.}
\label{fig:Frep-direction}
\end{center}
\end{figure}
%%%%%%%%%%%%%%%%%%%%%%%%%%%%%
The repulsive force then reads (see Fig.~\ref{fig:Frep-direction})
\begin{equation}
  \overrightarrow{F_{ij}^{\rm rep}} = -m_i k_{ij}
  \frac{v_{ij}^2}{R_{ij}}\overrightarrow{e_{ij}}\,.
\label{eq:CFMfrep}
\end{equation}
This definition of the repulsive force in the CFM reflects several aspects. 
First, the force between two pedestrians decreases with increasing 
distance. In the CFM it is inversely proportional to their distance $R_{ij}$.
Furthermore, the repulsive force takes into account the relative
velocity $v_{ij}$ between pedestrian $i$ and pedestrian $j$. The following
special definition provides that slower pedestrians are not affected
by the presence of faster pedestrians in front of them:
\begin{eqnarray}
  v_{ij} &=&
  \frac{1}{2}[(\overrightarrow{v_i}-\overrightarrow{v_j})\cdot
  \overrightarrow{e_{ij}} +
  |(\overrightarrow{v_i}-\overrightarrow{v_j})
  \cdot\overrightarrow{e_{ij}}|] \nonumber \\ 
 &=&\begin{cases}
   (\overrightarrow{v_i}-\overrightarrow{v_j})\cdot
   \overrightarrow{e_{ij}} & \quad\mbox{if}\;\;
   (\overrightarrow{v_i}-\overrightarrow{v_j})\cdot
   \overrightarrow{e_{ij}} > 0 \\ 0 & \quad\mbox{otherwise.}
 \end{cases}
\end{eqnarray}
As in general pedestrians  react only to obstacles and pedestrians that are
within their perception, the reaction field of the repulsive force is
reduced to the angle of vision ($180^\circ$) of each pedestrian, by
introducing the coefficient 
\begin{eqnarray}
  \label{eq:kij}
  k_{ij} & = &
  \frac{1}{2}\frac{\overrightarrow{v_i}\cdot\overrightarrow{e_{ij}} +
    \mid \overrightarrow{v_i}\cdot\overrightarrow{e_{ij}} \mid} {v_i}
  \nonumber \\ 
  & = &
\begin{cases}
  (\overrightarrow{v_i}\cdot\overrightarrow{e_{ij}})/v_i & \quad\mbox{if}\;\; 
   \overrightarrow{v_i}\cdot\overrightarrow{e_{ij}}>0\; \And\; 
    v_i \ne 0\\ 
  0 & \quad\mbox{otherwise.}
 \end{cases}
\end{eqnarray}
% \ne 0
The coefficient $k_{ij}$ is maximal when pedestrian $j$ is in the
direction of movement of pedestrian $i$ and minimal when the angle
between $j$ and $i$ is bigger than $90^\circ$. Thus the strength of
the repulsive force depends on the angle.

\begin{figure}
\begin{center}
\includegraphics[width=0.85\columnwidth]{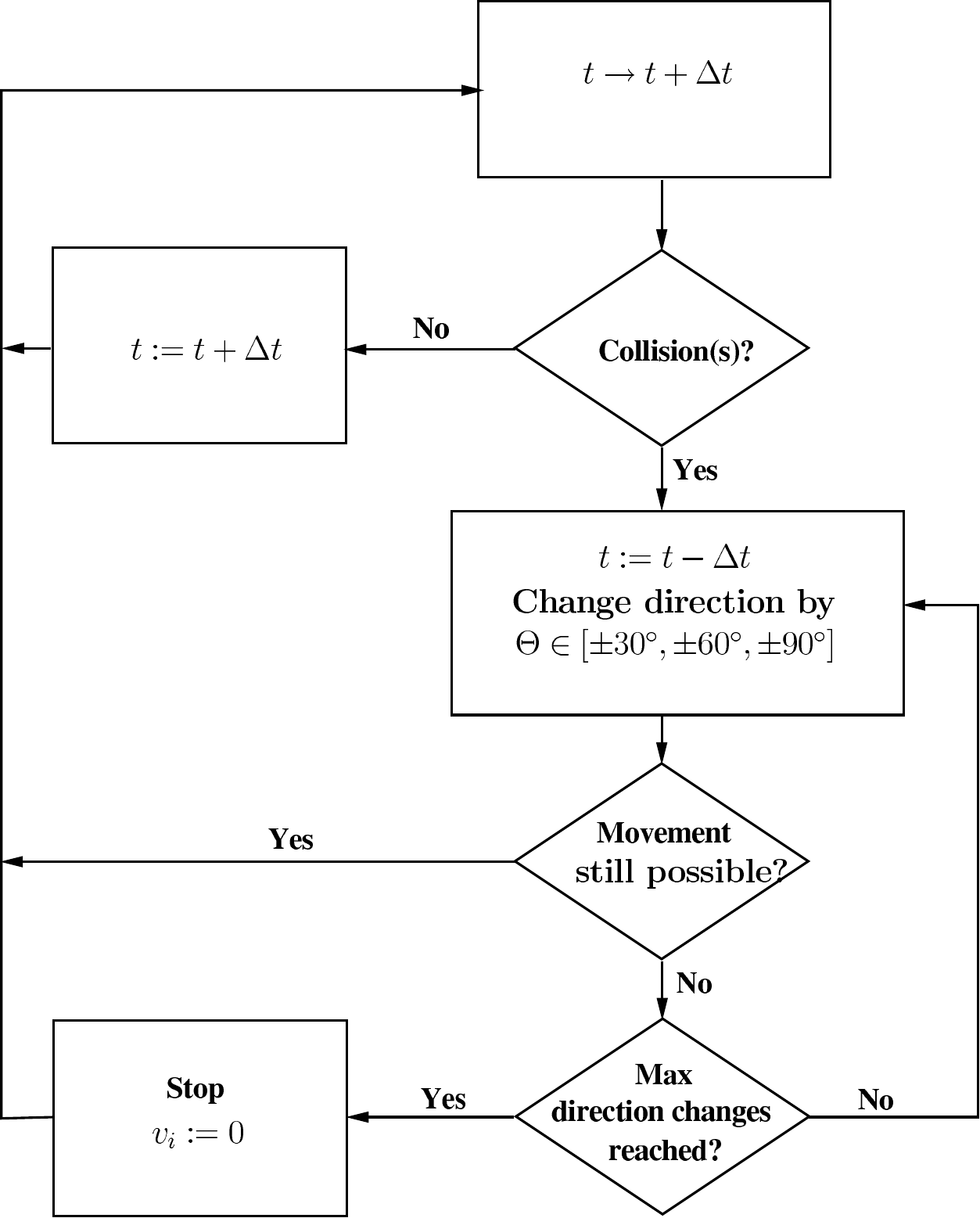}
\caption{Schematic representation of the collision detection technique
  (CDT), which is an important component in the CFM  \cite{Yu2005}, to manage collisions and mitigate overlapping among pedestrians. In our model we do not need the CDT, which is a considerable simplification in comparison to the CFM  \cite{Yu2005}.}
\label{fig:cdt}
\end{center}
\end{figure}

As mentioned earlier the CFM is complemented with a ``Collision
Detection Technique'' (CDT) to manage conflicts and mitigate
overlappings between pedestrians. Fig.~\ref{fig:cdt} depicts
schematically the definition of the CDT.
%%%%%%%%%%%%%%%%%%%%%%%%%%%%%%%%%%
%%%%%%%%%%%%%%%%%%%%%%%%%%%%%%%%%%
Although CDT is relatively simple, it adds an amount of complexity to
the initial model defined with Eq.~(\ref{eq:maineq}) and masks the
main idea behind the repulsive forces.
In the following we systematically modify the expression of the
repulsive force to enable a better quantitative description of
pedestrian dynamics.

%%%%%%%%%%%%%%%%%%%%%%%%%%%%%%%%%%%%%%%%%%%%%%%%%%%%%%%%%%%%%%%%%%%%%%%%%%

\section{Overlapping vs.\ Oscillation}

In this work we consider a velocity-dependent volume exclusion of
pedestrians.  Overlapping between two pedestrians occurs when their
geometrical form (circle, ellipse, ...) overlaps. Modelling a
pedestrian as a circle or ellipse is just an approximation of the
human body. Therefore, a certain amount of overlapping could be
acceptable and might be interpreted as ``elastic deformation''.
However, for the deformed body the center of mass no longer coincides
with the center of the circle or ellipse.  For this reason overlapping
is a serious problem that should be dealt with.

In~\cite{Chraibi2009a} it was shown that the introduction of a CDT is
necessary to mitigate overlapping among pedestrians. The CDT keeps
pedestrians away from each other with a distance of at least $r$,
where $r$ represents the radius of the circle modelling the volume
exclusion of pedestrians.

Our goal is to simplify the model by dispensing with the CDT and
improve the repulsive force to compensate for the effects of the missing
CDT on the dynamics.
To introduce the shape of the modeled pedestrians in 
Eq.~(\ref{eq:CFMfrep}) we transform the singularity of the repulsive
force from $0$ to $2r$:
\begin{equation}
  \overrightarrow{F_{ij}^{\rm rep}} = -m_i k_{ij}
  \frac{v_{ij}^2}{R_{ij} -2 r}\overrightarrow{e_{ij}}\,.
\label{eq:modCFMfrep1}
\end{equation}
Due to the quotient in Eq.~(\ref{eq:modCFMfrep1}) when the distance is
small, low relative velocities lead to an unacceptably small force.
Consequently, partial or total overlapping is not prevented.
Introducing the intended speed in the numerator of the repulsive force
eliminates this side-effect.  This dependence on the desired speed is
motivated by the observation that for faster pedestrians stronger
repulsive forces are required to avoid collisions with other
pedestrians and obstacles. Thus, the repulsive force is changed to
\begin{equation}
  \overrightarrow{F_{ij}^{\rm rep}}=-m_i k_{ij}\frac{(\eta v_i^0 +
    v_{ij})^2}{R_{ij} - 2r} \overrightarrow{e_{ij}},
\label{eq:modCFMfrep2}
\end{equation}
with a free parameter $\eta$ to adjust the strength of the force.
 
Those two changes in the repulsive force cause the emergence of two
phenomena: Overlapping and oscillations. In the following we will
define quantities to study those phenomena.

Avoiding overlapping between pedestrians and oscillations in their
trajectories is difficult to accomplish in force-based models. On one
hand, increasing the strength of the repulsive force with the aim of
excluding overlapping during simulations leads to oscillations in the
trajectories of pedestrians. Consequently backward movements occur,
which is not realistic especially in evacuation scenarios.

On the other hand, reducing the strength of the repulsive force (to
avoid oscillations) leads inevitably to overlapping between
pedestrians or between pedestrians and obstacles.
                       
To solve this dilemma one has to find an adequate value of the
strength of the repulsive force: it should neither be too high so that
oscillations will appear, nor too low so that overlapping will be
observed.

\begin{figure}
\begin{center}
\includegraphics[width=0.6\columnwidth]{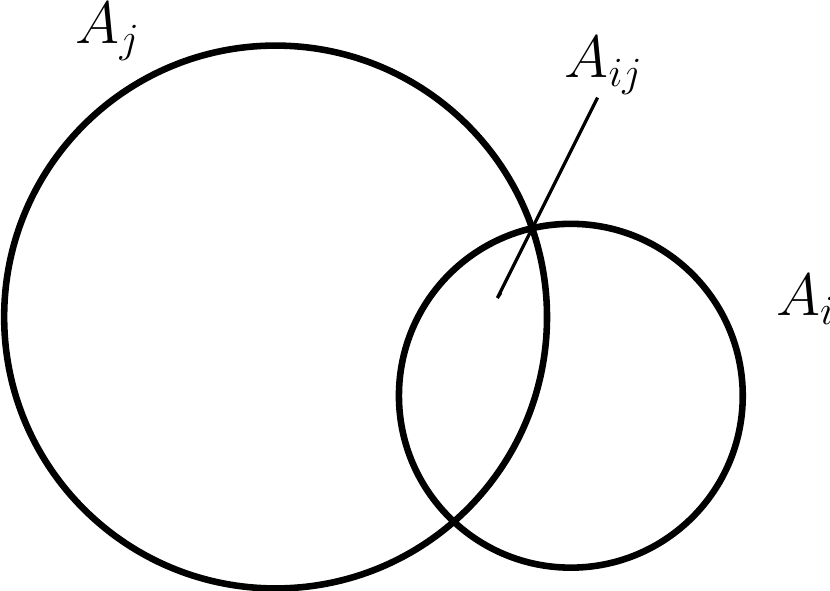}
\caption{The overlapping area between pedestrians $i$ and $j$ 
  varies between 0 and 1.}
\label{fig:Oij}
\end{center}
\end{figure}

To understand this duality we quantify overlapping and oscillations
during simulations.  First, we define an overlapping-proportion during
a simulation as:
\begin{equation}
  o^{(v)}= \frac{1}{n_{ov}}
  \sum_{t=0}^{t=t_\text{end}}\sum_{i=1}^{i=N}\sum_{j>i}^{j=N}o_{ij}\,,
\label{eq:ov}
\end{equation}
with
\begin{equation}
  o_{ij}=\frac{A_{ij}}{\min(A_i, A_j)}\, \le 1,
\label{eq:Oij}
\end{equation}
where $N$ is the number of simulated pedestrians. $A_{ij}$ is the
overlapping area of the circles $i$ and $j$ with areas $A_i$ and
$A_j$, respectively  (see Fig.~\ref{fig:Oij}).  
$n_{ov}$ is the cardinality of the set
\begin{equation}
\mathcal{O}:=\{o_{ij}: o_{ij} \ne 0 \}\,.
\end{equation}
For $n_{ov}=0$,\, $o^{(v)}$ is set to zero. 

%%%%%%%%%%%%%%%%%%%%%%%%%%
%%%%%%%%%%%%%%%%%%%%%%%%%%

For a pedestrian with velocity $\overrightarrow{v_i}$ and desired
velocity $\overrightarrow{v_{i}^0}$ we define the
oscillation-proportion as
\begin{equation}
  o^{(s)}= \frac{1}{n_{os}}\sum_{t=0}^{t=t_\text{end}}\sum_{i=1}^{i=N}S_{i}\,,
\label{eq:os}
\end{equation}
where $S_{i}$ quantifies the oscillation-strength of pedestrian $i$
and is defined as follows:
\begin{equation}
  S_{i}= \frac{1}{2} (-s_i + |s_i|)\,,
\label{eq:Si}
\end{equation}
with
\begin{equation}
s_{i}=\frac{\overrightarrow{v_i}\cdot\overrightarrow{v_{i}}^0}{\left(
v_{i}^0\right)^2}\,,
\label{eq:si}
\end{equation}
and $n_{os}$ is the cardinality of the set
\begin{equation}
\mathcal{S}:=\{s_{i}: s_{i} \ne 0 \}.
\end{equation}
Here again $o^{(s)}$ is set to zero if $n_{os}=0$.  The proportions $o^{(v)}$
and $o^{(s)}$ are normalized to 1 and describe the evolution of the
phenomena overlapping and oscillations during a simulation.

\begin{figure}
\begin{center}
\includegraphics[width=1\columnwidth]{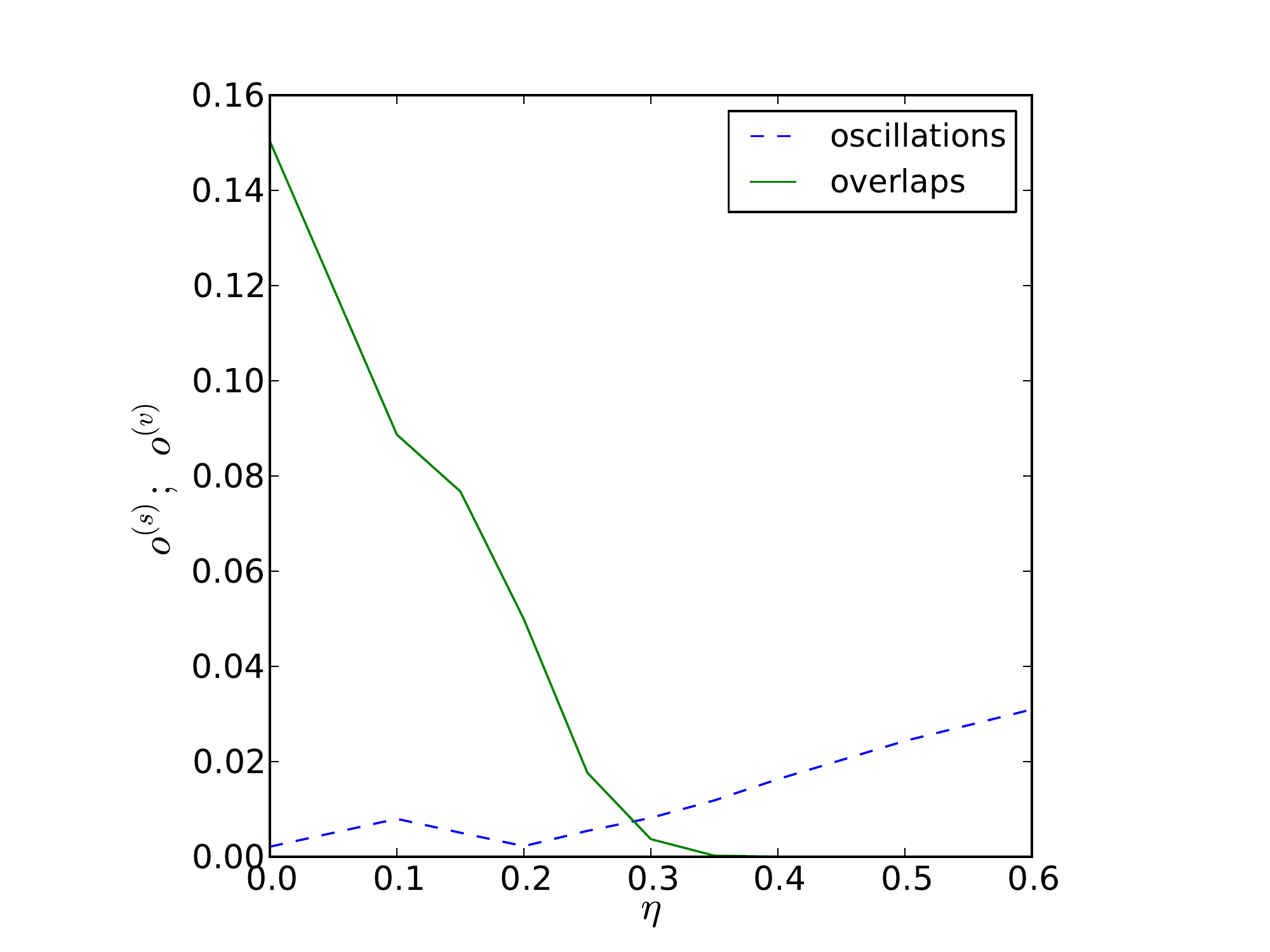}
\caption{(Color online) Oscillation-proportion $o^{(s)}$ and overlapping-proportion
$o^{(v)}$ as function of the interaction strength $\eta$ obtained from
200 simulations with different initial conditions.
  Oscillations increase with increasing strength of the
repulsive force, while overlaps become negligible for larger $\eta$. The case $\eta=0$
  is the CFM. In each run the simulations for different $\eta$ are
  started with the same initial values.
 }
\label{fig:os-ov}
\end{center}
\end{figure}

In order to exemplify the behavior of these two coupled phenomena we
simulate an evacuation of 35 pedestrian from a $4\, \text{m}\times 4\,
\text{m}$ room with an exit of $1.2\, \text{m}$ and determine
$o^{(v)}$ and $o^{(s)}$ for different values of $\eta$ in
Eq.~(\ref{eq:modCFMfrep2}).  Results are shown in Fig.~\ref{fig:os-ov}. 
$\eta=0$ is a special case of the model and represents the CFM  \cite{Yu2005}. The high values of the overlapping 
  proportion suggest that simulations using only CFM 
  without the CDT lead to unreasonable results. For further
  details we refer to~\cite{Chraibi2009a}.
 
The introduction of the intended velocity in the repulsive force
enhances the ability of the repulsive force to guarantee the volume
exclusion of pedestrians. This is reflected by the decreasing of the
overlapping-proportion $o^{(v)}$ while increasing $\eta$
(Eq.~\ref{eq:modCFMfrep2}). See Fig.~\ref{fig:os-ov}.

Meanwhile, the oscillation-proportion $o^{(s)}$ increases, thus the
system tends to become instable.  Large values of the
oscillation-proportion $o^{(s)}$ imply less stability. For $s_i=1$ one
has $\overrightarrow{v_i}=- \overrightarrow{v_{i}}^0$, i.e.\ a
pedestrian moves backwards with desired velocity. Even values of $s_i$
higher than 1 are not excluded and can occur during a simulation.
Therefore, a careful calibration of $\eta$ is required to achieve an
optimal balance between overlapping and oscillations.
  
Unfortunately, it is not possible to adjust the strength of the
repulsive force by means of $\eta$ in order to get an overlapping-free
and meanwhile an oscillation-free simulation. Nevertheless, by proper choice of $\eta$ one can reduce the amount of overlapping among pedestrians such that it becomes negligible and can
be interpreted as a deformation. This characteristic of the GCFM is
not fulfilled by the CFM \cite{Yu2005}, where total overlapping ($o_{ij}=1$) can be observed. 

Furthermore, the
quantities $o^{(s)}$ and $o^{(v)}$ provide a criterion to choose an
optimal value for $\eta$, which is given by the intersection of the
curves representing $o^{(s)}$ and $o^{(v)}$.

%%%%%%%%%%%%%%%%%%%%%%%%%%%%%%%%%%%%%%%%%%%%%%%%%%%%%%%%%%%%%%%%%%%%%%%%%%

\section{Hard circles vs. Dynamical circles: The fundamental diagram 
for single file movement}

It is suggested that the effective space requirement of a
moving pedestrian varies with velocity.  Usually, the projection of
the pedestrian's shape to the two-dimensional plane is modeled as a
circle with a radius $r$ \cite{Helbing1995,Johansson2007,Parisi2007}.
Thompson suggested a three-circle representation for main body and
shoulders \cite{Thompson1995}. According to \cite{Pauls2004}, however,
the radius of the circle varies such that the space requirement of
pedestrians increases significantly as speed increases. In \cite{Seyfried2006} a linear velocity-dependence
\begin{equation}
r_i= r_\text{min} + \tau_r v_i
\label{eq:diam}
\end{equation}
of the radius with parameters $r_\text{min}$\, and $\tau_r$ was
suggested. ``Space requirement'' encompasses the physical area taken by the torso together with the motion of the legs, lateral swaying and a safety margin.

The repulsive force reads
\begin{equation}
   \overrightarrow{F_{ij}^{\rm rep}}=-m_i k_{ij}\frac{(\eta {v_i}^0 +
     v_{ij})^2}{\operatorname{dist}_{ij}} \overrightarrow{e_{ij}},
\label{eq:modCFMfrep3}
\end{equation}
with 
\begin{equation}
\operatorname{dist}_{ij} = R_{ij} - r_i(v_i) -  r_j(v_j)  
\end{equation}
the effective distance between pedestrian $i$ and $j$ and $r_i$ the
radius of pedestrian $i$ as defined in Eq.~(\ref{eq:diam}).

%%%%%%%%%%%%%%%%%%%%%%%%%%%%%%%%%%%%%%%%%%%%%%%%%%%%%%%%%%%%%%%%%%%

\section{Elliptical Volume Exclusion of Pedestrians}

One drawback of circles that impact negatively the dynamics is their
rotational symmetry with respect to their centers. Therefore, they
occupy the same amount of space in all directions. In single
file movement this is irrelevant since the circles are projected
to lines and only the required space in movement direction matters.
However, in two dimensional movement the aforementioned symmetry lasts
by occupying unnecessary lateral space.

In \cite{Fruin1971} Fruin introduced the ``body ellipse'' to describe
the plane view of the average adult male human body. Pauls
\cite{Pauls2004} presented ideas about an extension of Fruin's ellipse
model to better understand and model pedestrian movement as density
increases. Templer \cite{Templer1992} noticed that the so called
``sensory zone'', which is a bubble of space between pedestrians and
other objects in the environment to avoid physical conflicts and for
psychocultural reasons, varies in size and takes the shape of an
ellipse. In fact, ellipses are closer to the projection of required space of the human body on the plane,
including the extent of the legs during motion and the lateral swaying of the body.

Having the ambition to describe with the same set of parameters the
dynamics in one- and two-dimensional space we extend our model by
introducing an elliptical volume exclusion of pedestrians.
Given a pedestrian $i$ we define an ellipse with center ($x_i$,$y_i$),
major semi-axis $a$ and minor semi-axis\, $b$.  $a$ models the space
requirement in the direction of movement. In analogy to
Eq.~(\ref{eq:diam}) we set
\begin{equation}
a= a_\text{min} + \tau_a v_i
\label{eq:a}
\end{equation}
with two parameters $a_\text{min}$\, and $\tau_{a}$.

\begin{figure}
 \begin{center}
  \subfigure{\label{fig:swayslow}}
  \includegraphics[width=0.48\columnwidth]{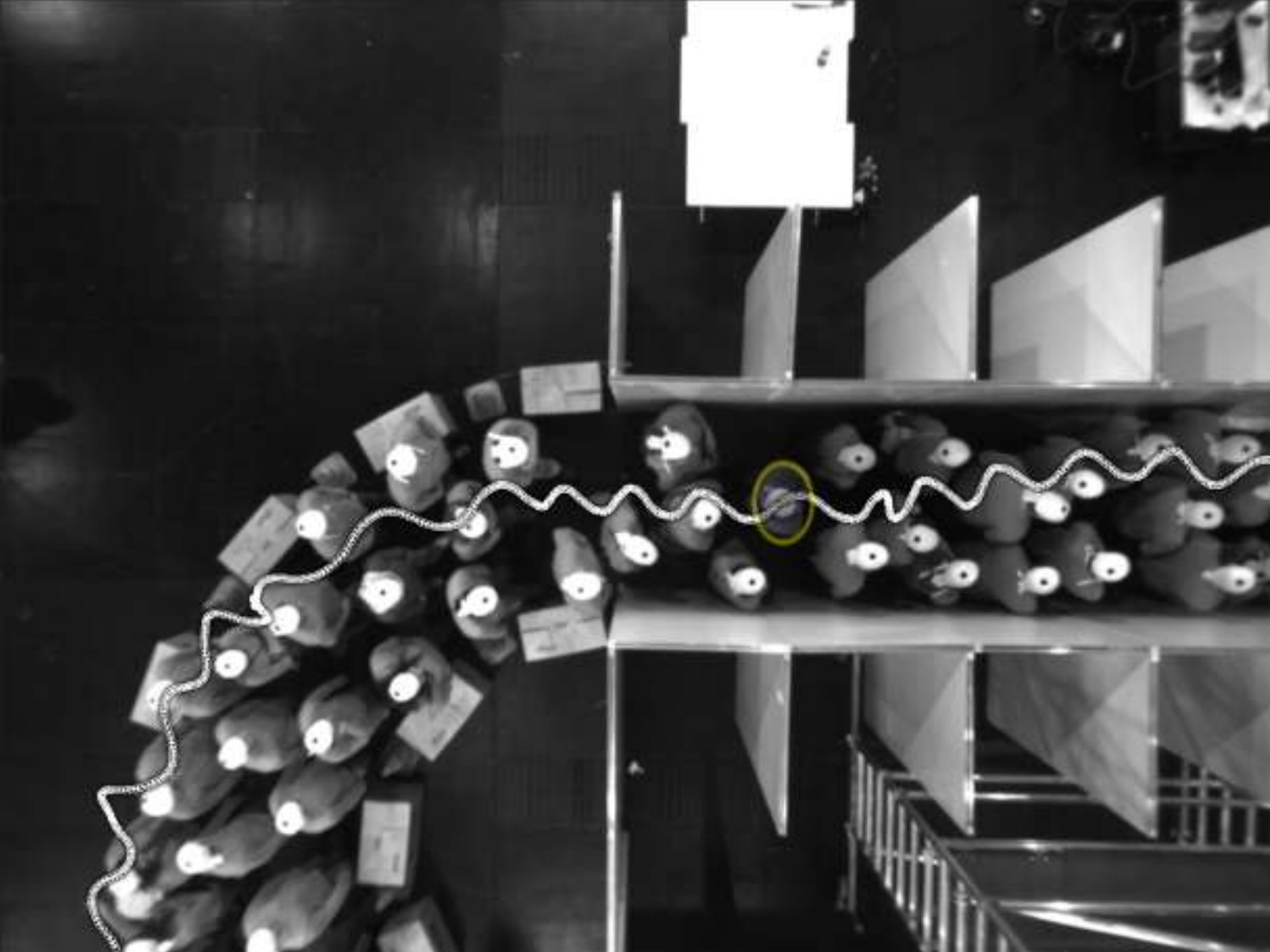}
  \subfigure{\label{fig:swayfast}}
  \includegraphics[width=0.48\columnwidth]{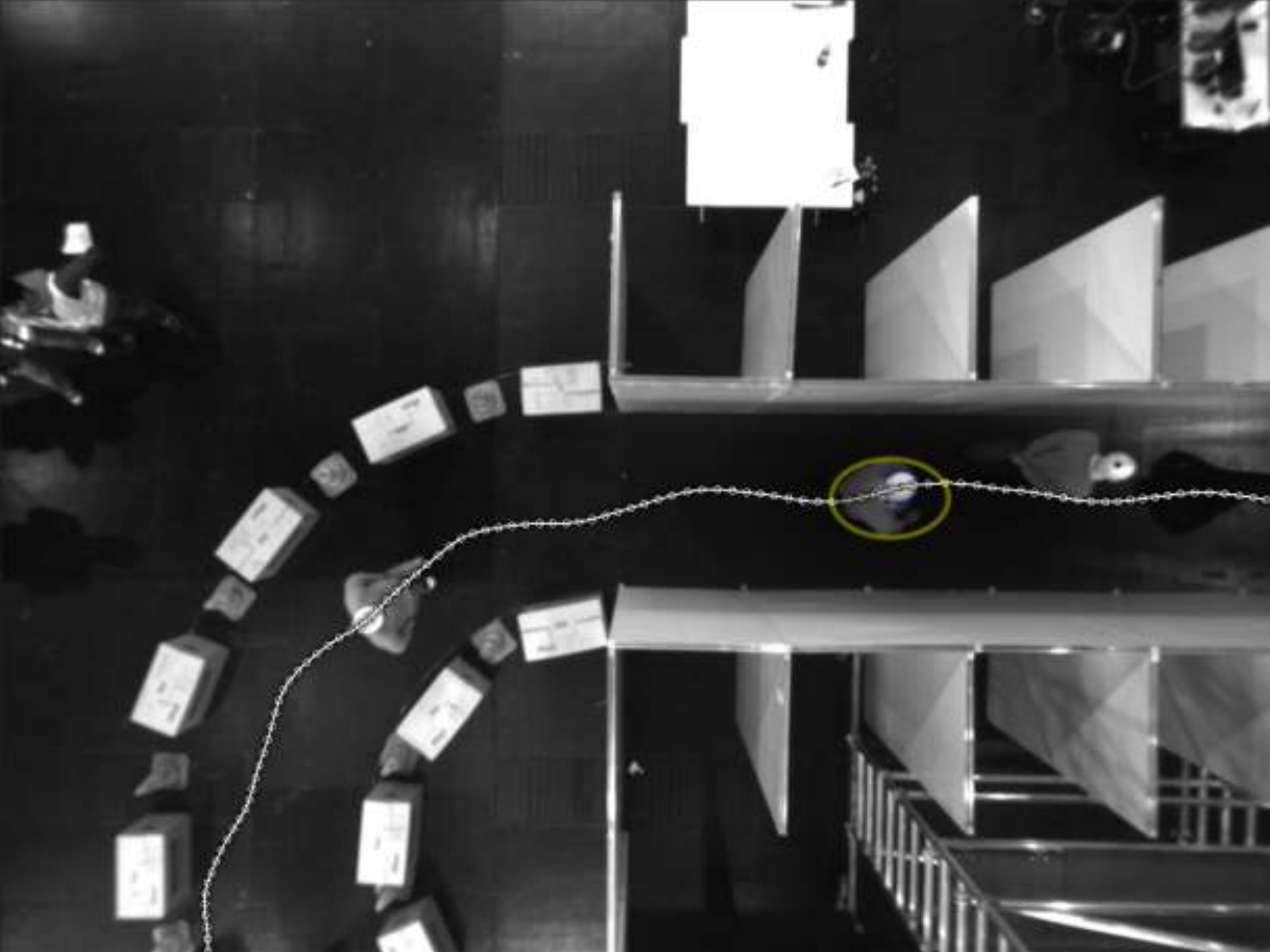}
  \caption{Off-line trajectory detection with PeTrack \cite{Boltes2009}. 
    Left: The trajectory of the detected pedestrian shows
    strong swaying. Right: The faster pedestrians move, the smoother
    and weaker is the swaying of their trajectories.
}
  \label{fig:sway}
 \end{center}
\end{figure}

Fruin \cite{Fruin1971} observed body swaying during both human
locomotion and while standing. Pauls \cite{Pauls2006b} remarks that
swaying laterally should be considered while determining the required
width of exit stairways.  In \cite{Hoogendoorn2005} characteristics of
lateral swaying are determined experimentally.
Observations of experimental trajectories in \cite{Hoogendoorn2005}
indicate that the amplitude of lateral swaying varies from a maximum
$b_\text{max}$ for slow movement and gradually decreases to a minimum
$b_\text{min}$ for free movement when pedestrians move with their free
velocity (Fig.~\ref{fig:sway}).
Thus we describe with $b$ the lateral swaying of pedestrians and set
\begin{equation}
  b= b_\text{max} - (b_\text{max}-b_\text{min})\frac{v_i}{v_{i}^0}
\label{eq:b}
\end{equation}
Since $a$ and $b$
are velocity-dependent, the inequality 
\begin{equation}
b \le  a
\end{equation} 
does not always hold for the ellipse $i$. In the rest of this work we denote
the semi-axis in the movement direction by $a$ and its
orthogonal semi-axis by $b$.

%%%%%%%%%%%%%%%%%%%%%%%%%%%%%%%%%%%%%

%%%%%%%%%%%%%%%%%%%%%%%%%%%%%%%%%%%%%%%%%%%%%%%%%%%%%%%%%%%%%%%%%%%%%%%%%%
\section{Elliptical Volume Exclusion and Force Implementation}

In this section we give some mathematical insights concerning the
implementation of the repulsive forces.

%%%%%%%%%%%%%%%%%%%%%%%%%%%%%%%%%%%%%

\subsection{Repulsive Forces between Pedestrians}
\label{sec:rep-for}
In order to calculate the repulsive force emerging from pedestrian $j$ acting on pedestrian $i$ according to Eq.~(\ref{eq:modCFMfrep3})
we require the distance between the borders of the ellipses, along a
line connecting the two pedestrians $\operatorname{dist}_{ij}$. See App.~\ref{sec:dist} for more details on $\operatorname{dist}_{ij}$.

Another important quantity is the distance of closest approach or
contact distance of two ellipses $\tilde{l}$ which is the minimum of $\operatorname{dist}_{ij}$
while $i$ and $j$ are not
overlapping. Unlike for circles, $\tilde{l}$ can be non-zero for
ellipses and depends on their orientations. In \cite{Zheng2007} an
analytical expression for the distance of the closest approach of two
ellipses with arbitrary orientation is derived. Fig.~\ref{fig:interp-dir} shows how  $\operatorname{dist}_{ij}$ and $\tilde{l}$ goes in the repulsive force.

\subsection{Repulsive Forces between Pedestrians and Walls}

\begin{figure}
 \begin{center}
  \includegraphics[width=0.6\columnwidth]{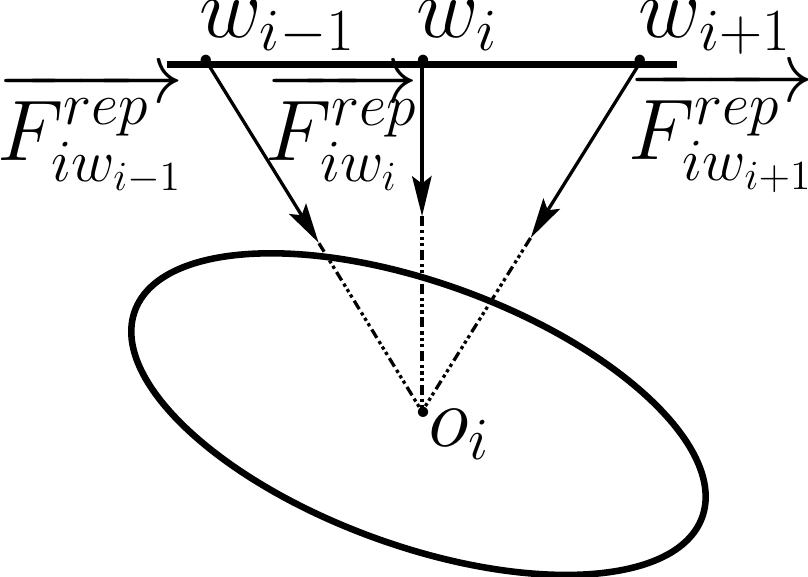}
  \caption{Each wall is modelled as three static point masses 
    acting on pedestrians.}  
  \label{fig:fuss}
 \end{center}
\end{figure}

The repulsive force between a pedestrian $i$ and a wall is zero if $i$
performs a parallel motion to the wall.  While this behavior of the
force is correct, it leads to very small repulsive forces when the
pedestrians motion is almost parallel to the wall.  For this reason we
characterize in this model walls by three point masses 
acting on pedestrians within a certain interaction range (Fig.~\ref{fig:fuss}).
The middle point is the point with the shortest distance from the
center of the pedestrian to the line segment of the wall. All three
points have to be computed at each step as the
pedestrian moves. The distance between the three wall points is set
to the minor semi-axis of an ellipse 
If one lateral point ($w_{i+1}$ or $w_{i+1}$ ) does not lie on the
line segment of the wall, then it will not be considered in the
computation of the repulsive force.

The number of point masses have been chosen by a
process of trial and error.  Simulations have shown that three point
masses are sufficient to keep pedestrians away from walls.  Meanwhile
they are computationally cost-effective.

As walls are static objects, the repulsive force emerging from a wall
$w$ and acting on pedestrian $i$ simplifies to
\begin{equation}
  \overrightarrow{F_{iw}^{\rm rep}}=\sum_{j=i-1 }^{i+1}
  \overrightarrow{F_{iw_{j}}^{\rm rep}},
\end{equation}
with 
\begin{equation}
  \overrightarrow{F_{iw_j}^{\rm rep}}=-m_i k_{iw_i}\frac{(\eta v_i^0 +
    v^n_{i})^2}{\operatorname{dist}_{iw_j}} \overrightarrow{e_{iw_j}},
  \qquad j \in  \{i-1,\, i,\, i+1 \}\, .
\label{eq:modCFMfrepw}
\end{equation}
$v^n_{i}$ is the component of the velocity normal to the wall,
$k_{iw_i}$ and $\overrightarrow{e_{iw_j}}$ as defined resp. in
Eqs.~(\ref{eq:kij}) and (\ref{eq:R}) in Sec.~\ref{sec:CFM}.

%%%%%%%%%%%%%%%%%%%%%%%%%%%%%%%%%%%%%
%%%%%%%%%%%%%%%%%%%%%%%%%%%%%%%%%%%%%
The distance between a line $w$ and the ellipse $i$ is
\begin{equation}
\operatorname{dist}_{iw} = k_i - r_i,
\label{eq:distW}
\end{equation}
with $r_i$ the polar radius determined in Eq.~(\ref{eq:quad}) and
$k_i$ the distance of point $o_i$ to the line $w$. Further details
can be found in App.~\ref{sec:dist-wall}. According to the distance $\tilde{l}$ defined for the repulsive
forces between pedestrian in Sec.~\ref{sec:rep-for} we introduce the distance of
the closest approach between an ellipse and a line $\tilde{k}$, see App.~\ref{sec:dist-wall} and Fig.~\ref{fig:interp-dir} for details.

Note that in Eq.~(\ref{eq:modCFMfrepw}), $k_{iw_i}$ in the force is
independent of the chosen lateral wall point $w_j$. That means, if a
pedestrian is moving parallel to the wall, $k_{iw_i}=0$ and thus the
points $j-1$\, and $j+1$ have no effects.

%%%%%%%%%%%%%%%%%%%%%%%%%%%%%%%%%%%%%%%%%%%%%%%%%%%%%%%%%%%%%%%%%%%%%%%%%%

\subsection{Numerical Stabilization of the Repulsive Force}

In this section we describe a numerical treatment of the repulsive
force. For the sake of simplicity, we focus on the case of pedestrian-pedestrian
interactions. The pedestrian-wall case is treated similarly.

\begin{figure}
 \begin{center}
  \includegraphics[width=0.75\columnwidth]{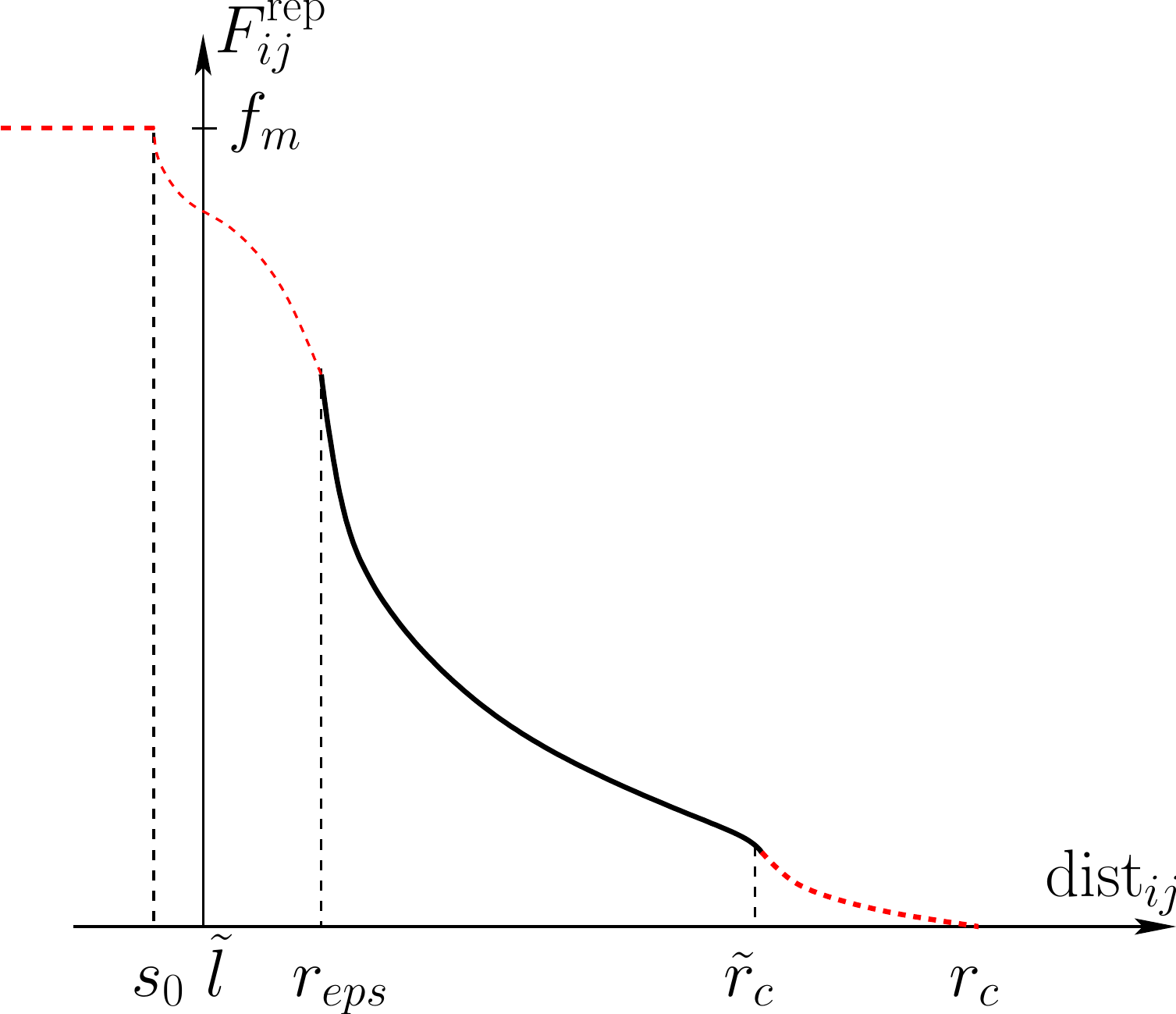}
  \caption{(Color online) The interpolation of the repulsive 
    force between pedestrians $i$ and $j$ Eq.~(\ref{eq:modCFMfrep3}) depending on  $\operatorname{dist}_{ij}$ and the distance of closest approach $\tilde{l}$, see Sec.~\ref{sec:rep-for}. As the repulsive
    force also depends on the relative velocity $v_{ij}$, this figure
    depicts the curve of the force for $v_{ij} = \mbox{const}$. The left and right dashed curves are defined in
      Eqs.~(\ref{eq:Pl}) and~(\ref{eq:Pr}) respectively. The wall-pedestrian interaction has an analogous form with $\operatorname{dist}_{ij}$
and $\tilde{l}$ replaced by $\operatorname{dist}_{wi}$ and $\tilde{k}$, respectively.
}
  \label{fig:interp-dir}
 \end{center}
\end{figure}

The strength of the repulsive force decreases with increasing distance
between two pedestrians. Nevertheless the range of the repulsive force
is infinite. This is unrealistic for interactions between pedestrians.
Therefore, we introduce a cut-off radius $r_c=2~\text{m}$ for the
force limiting the interactions to adjacent pedestrians solely. To
guarantee robust numerical integration a two-sided
Hermite-interpolation of the repulsive force is implemented. The
interpolation guarantees that the norm of the repulsive force
decreases smoothly to zero for $\operatorname{dist}_{ij} \rightarrow
r^-_{c}$. For
$\operatorname{dist}_{ij} \rightarrow \tilde{l}^+$ 
the interpolation avoids an increase of the force to infinity but to
$f_m = 3 F_{ij}^\text{rep}(r_\text{eps})$ at $s_0=r_\text{eps}$ \, and $r_\text{eps}=0.1~\text{m}$, where
it remains constant. $\operatorname{dist}_{ij}$ and $\tilde{l}$ are illustrated in Sec.~\ref{sec:rep-for}. Fig.~\ref{fig:interp-dir} shows the dependence of the repulsive force on the distance for constant relative
velocity.

The right interpolation function $P_r$ and the left one $P_l$ (dashed
parts of the function in Fig.~\ref{fig:interp-dir}) are defined using
\begin{eqnarray}
   P_r(\tilde{r}_c)&&=F_{ij}^\text{rep}(\tilde{r}_c),\; P_r(r_c)=0
   \quad  \nonumber\\
   \quad (P_{r})^\prime(\tilde{r}_c)&&=\left(F_{ij}^\text{rep}\right)^\prime
   (\tilde{r}_c), 
   \;  (P_{r})^\prime(r_c)=0
\label{eq:Pr}
\end{eqnarray}
with $\tilde{r}_c=r_c-r_\text{eps}$ and 
\begin{eqnarray}
  P_{l}(s_0)&&=f_m, \;
  P_{l}(r_\text{eps})=F_{ij}^\text{rep}(r_\text{eps})
  \quad \nonumber\\
  (P_l)^\prime(s_0^+)&&=1,\;
  (P_l)^\prime(r_\text{eps})=\left(F_{ij}^\text{rep}\right)^\prime
  (r_\text{eps})\,.
\label{eq:Pl}
\end{eqnarray}
where the prime indicates the derivative.  $s_0$ is the minimum
allowed magnitude of the effective distance of two ellipses. Due to the 
superposition of the forces the inequality:
\begin{equation}
\operatorname{dist}_{ij}\ge s_0 \,.
\end{equation}
for pedestrians $i$ and $j$ is not guaranteed. 

%%%%%%%%%%%%%%%%%%%%%%%%%%%%%%%%%%%%%%%%%

\section{Simulation results}

The initial value problem in Eq.~(\ref{eq:maineq}) was solved using an
Euler scheme with fixed-step size $ \Delta t=0.01\; \text{s}$. First
the state variables of all pedestrians are determined. Then the update
to the next step is performed. Thus, the update in each step is
parallel.

The desired speeds of pedestrians are Gaussian distributed with mean
$\mu = 1.34~\text{m/s}$ and standard deviation $\sigma=
0.26~\text{m/s}$. The time constant $\tau$ in the driving force
Eq.~(\ref{eq:fdrv}) is set to $ 0.5~\text{s}$, i.e.\ $\tau\gg \Delta t$. 
For simplicity, the mass $m_i$ is set to unity.

In order to verify the model and evaluate the difference of the
elliptical shape of the volume exclusion versus the circular one we
measure the fundamental diagram in two-dimensional space with the same
set of parameter as for the one-dimensional fundamental diagram.  In
the one-dimensional case only the space requirement of pedestrians in
movement direction, expressed in terms of the semi-axis $a$, influences
the dynamics of the system. We set $a_\text{min}=0.18~\text{m}$ and
$\tau_{a}=0.53~\text{s}$ (see Eq.~\ref{eq:a}). 
%%%%%%%%%%%%%%%%%%%%%%%%%%%%%%%%%%%%
\begin{figure}
\begin{center}
 \subfigure{\label{fig:fd1d-emp}}
\includegraphics[width=0.75\columnwidth]{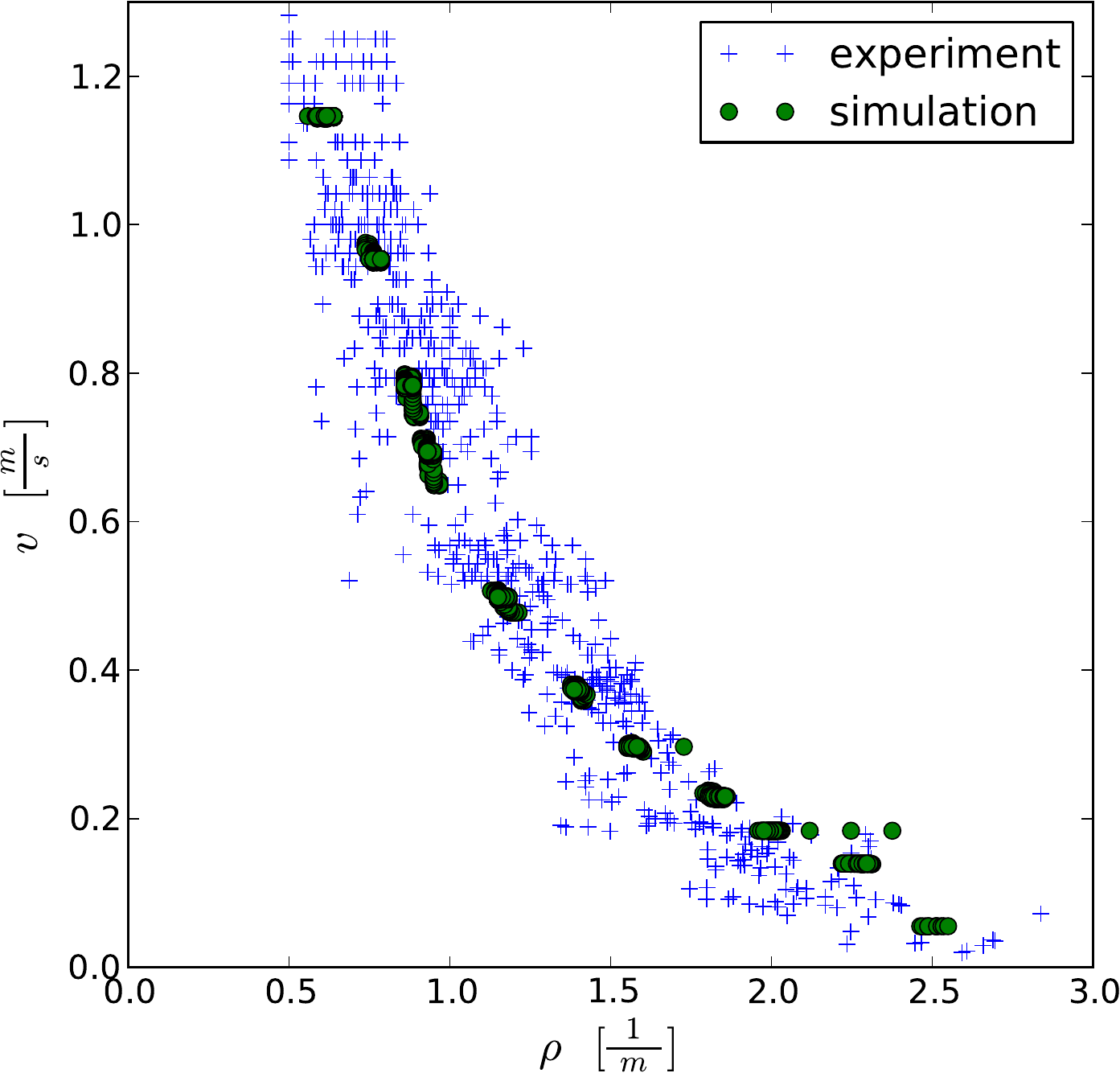}
\vspace{0.5cm}
\subfigure{\label{fig:fd1d-b}}
\includegraphics[width=0.75\columnwidth]{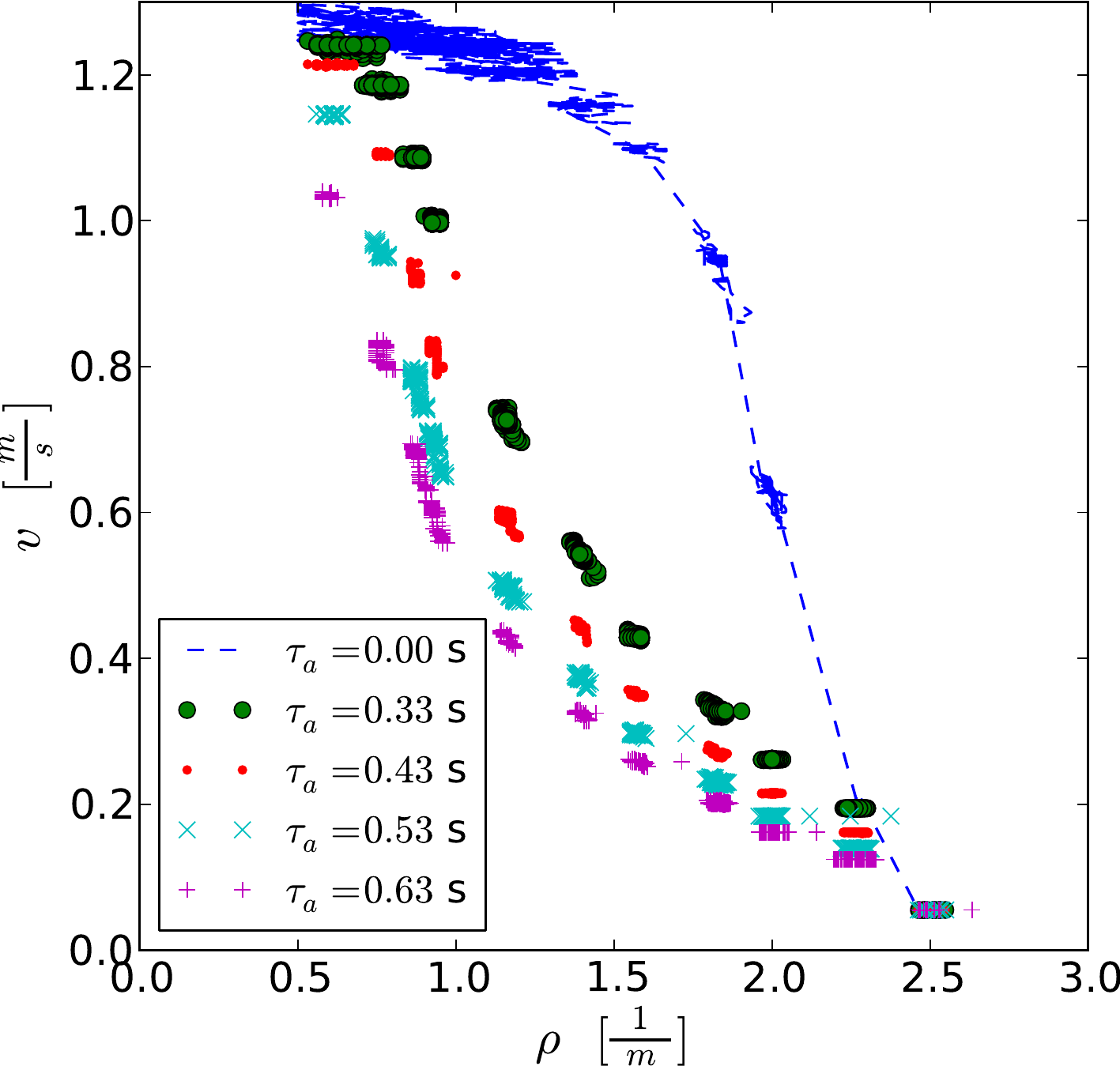}
\caption{(Color online) Top: Velocity-density relation for 
  one-dimensional movement compared to experimental data
  \cite{Seyfried2009a}. For the
      simulations, $\tau_{a}$ is set to $0.53~\text{s}$ . Bottom:
  Changing $\tau_{a}$ in Eq.~(\ref{eq:a}) influences the slope of the
  diagram.  $a_{min}$ has been kept equal to $0.18\,\text{m}$.
  $\tau_{a}=0$ represents pedestrians with constant
    space-requirement. }
\label{fig:fd1d-bk}
\end{center}
\end{figure}

%%%%%%%%%%%%%%%%%%%%%%%%%%%%%%%%%%%%
To illustrate the impact of the velocity-dependence of the radius on
the dynamics of pedestrians we measure the one-dimensional fundamental
diagram in a corridor of $26\, \text{m}$ with periodic boundary
conditions.  The measurement segment is 2~m long and situated in the
middle of the corridor.  Details about the measurement method are
given in App.~\ref{sec:meas}.
%%%%%%%%%%%%%%%%%%%%%%%%%%%%%%%%%%%%

The results for the one-dimensional fundamental diagram are shown
in Fig.~\ref{fig:fd1d-bk} and compare well with experimental data. Ellipses with velocity-dependent semi-axes emulate the space requirement of 
the projected shape of pedestrians better. Even the
shape of the fundamental diagram is reproduced after inclusion of this
velocity-dependence.

\begin{figure}
 \begin{center}
  \includegraphics[width=0.75\columnwidth]{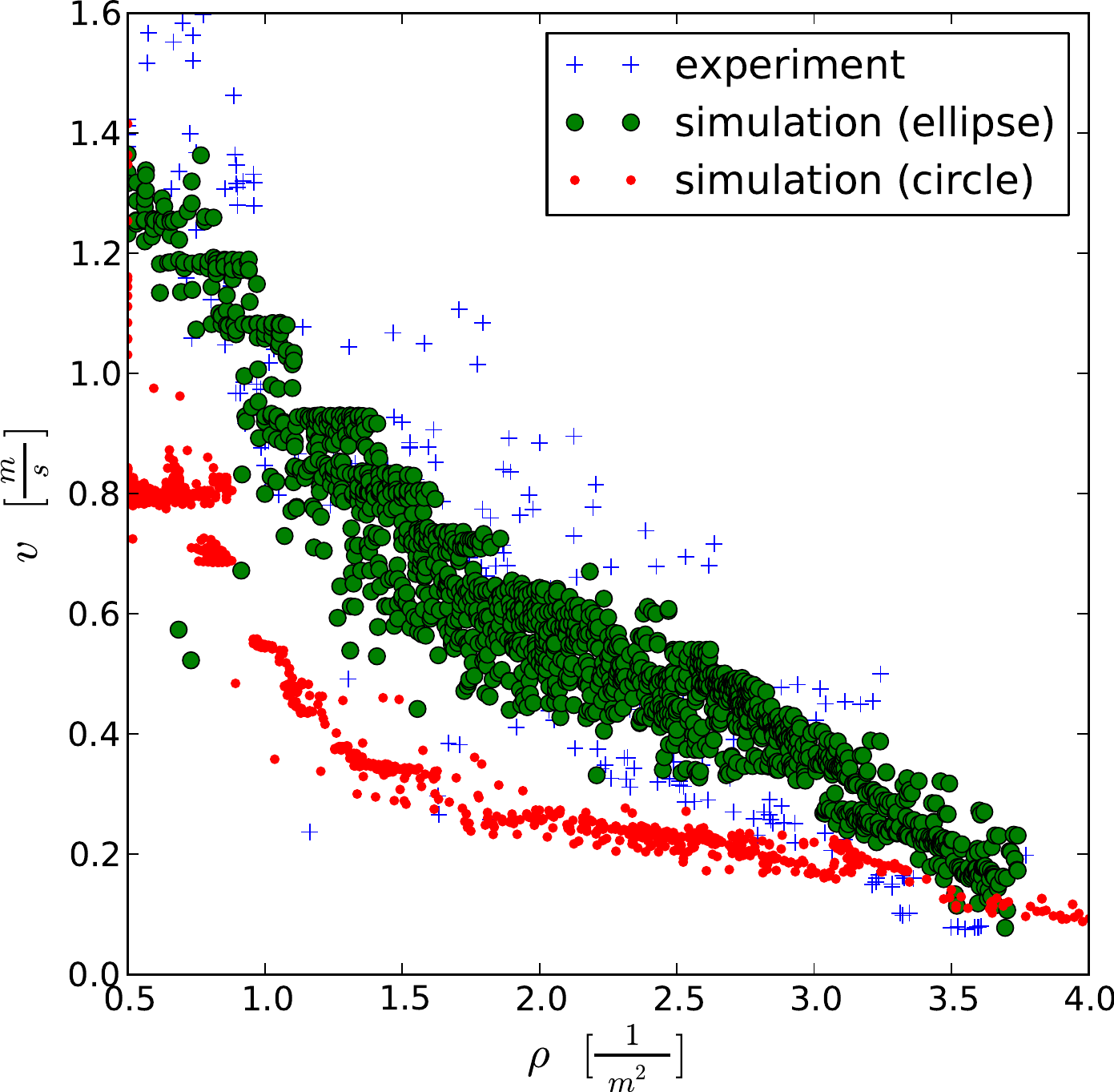}
  \caption{(Color online) Density-velocity relation in a corridor of dimensions 
    $25\, \text{m}\times1\, \text{m}$ in comparison with experimental
    data obtained in the HERMES-project \cite{Holl2009}. For the
    simulation with circles, $b$ is set to be equal to $a$.
}
  \label{fig:2dfd}
 \end{center}
\end{figure}

We extend the simulation to two-dimensional space and simulate a 
$25~\text{m}\times 1~\text{m}$ corridor with periodic boundary
conditions. A measurement segment of $2~\text{m}\times 1~\text{m}$ was
set in the middle of the corridor. We use the same measurement method
as for the single-file case (see App.~\ref{sec:meas}).  Calibration of
the parameters of the lateral semi-axis $b$ ($b_\text{min}$ and
$b_\text{max}$ in Eq.~\ref{eq:b}) leads to the values
$b_\text{min}=0.2\, \text{m}$ and $b_{max}=0.25~\text{m}$. The
simulation result is shown in Fig~\ref{fig:2dfd}.

%%%%%%%%%%%%%%%%%%%%%%%%%%%%%%%%%%%%
%%%%%%%%%%%%%%%%%%%%%%%%%%%%%%%%%%%%

With the chosen dimensions of the semi-axes $a$ and $b$ the model
yields the right relation between velocity and density both in
single-file movement and wide corridors, although only a corridor
width of 1~m was investigated. One remarks that the fundamental
diagram for elliptical shaped particles is an upper bound for that of
circular ones, especially at low and medium densities. At high
densities there is no noticeable difference between both shapes.

%%%%%%%%%%%%%%%%%%%%%%%%%%%%%%%%%%%%%%%%%%%%%%%%%%%%%%%%%%%%%%%%%%%%%%%%%%

\section{Conclusions}

We have proposed modifications of a spatially continuous force-based
model~\cite{Yu2005} to describe quantitatively the movement of
pedestrians in one- and two-dimensional space. Besides being a remedy
for numerical instabilities in CFM the modifications simplify the
approach of Yu et al.~\cite{Yu2005} since we can dispense with their
extra ``collision detection technique'' without deteriorating
performance. The implementation of the model is straightforward and
does not use any restrictions on the velocity. Furthermore, we
introduced an elliptical volume exclusion of pedestrians and studied
its influence compared to the standard circular one.  Simulation
results show good agreement with experimental data. Nevertheless, the
model contains free parameters that have to be tuned adequately to
adapt the model to a given scenario. Further improvement of the model
could be made by including, for example, a density-dependent repulsive
force.

Although the model describes quantitatively well the operative level
of human behavior, it does not consider aspects of the tactical and
strategic levels \cite{Schadschneider2009}. Phenomena like
cooperation, changing lanes and overtaking are not reproduced,
especially in bi-directional flow.

The source code of this model will be released under the GNU General Public Licence (GPL~\cite{GPL}) and will  be available  for download from  \cite{JSCPED}.
%%%%%%%%%%%%%%%%%%%%%%%%%%%%%%%%%%%%%%%%%%%%%%%%%%%%%%%%%%%%%%%%%%%%%%%%%%

\begin{acknowledgments}
The authors are grateful to the Deutsche Forschungsgemeinschaft
(DFG) for funding this project under Grant-Nr.: SE 1789/1-1.
\end{acknowledgments}

%%%%%%%%%%%%%%%%%%%%%%%%%%%%%%%%%%%%%%%%%%%%%%%%%%%%%%%%%%%%%%%%%%%%%%%%%%

%\begin{appendix} 
%\section{Appendix}
\appendix
\section{Distance between two ellipses}
\label{sec:dist}

\begin{figure}
\begin{center}
\includegraphics[width=0.7\columnwidth]{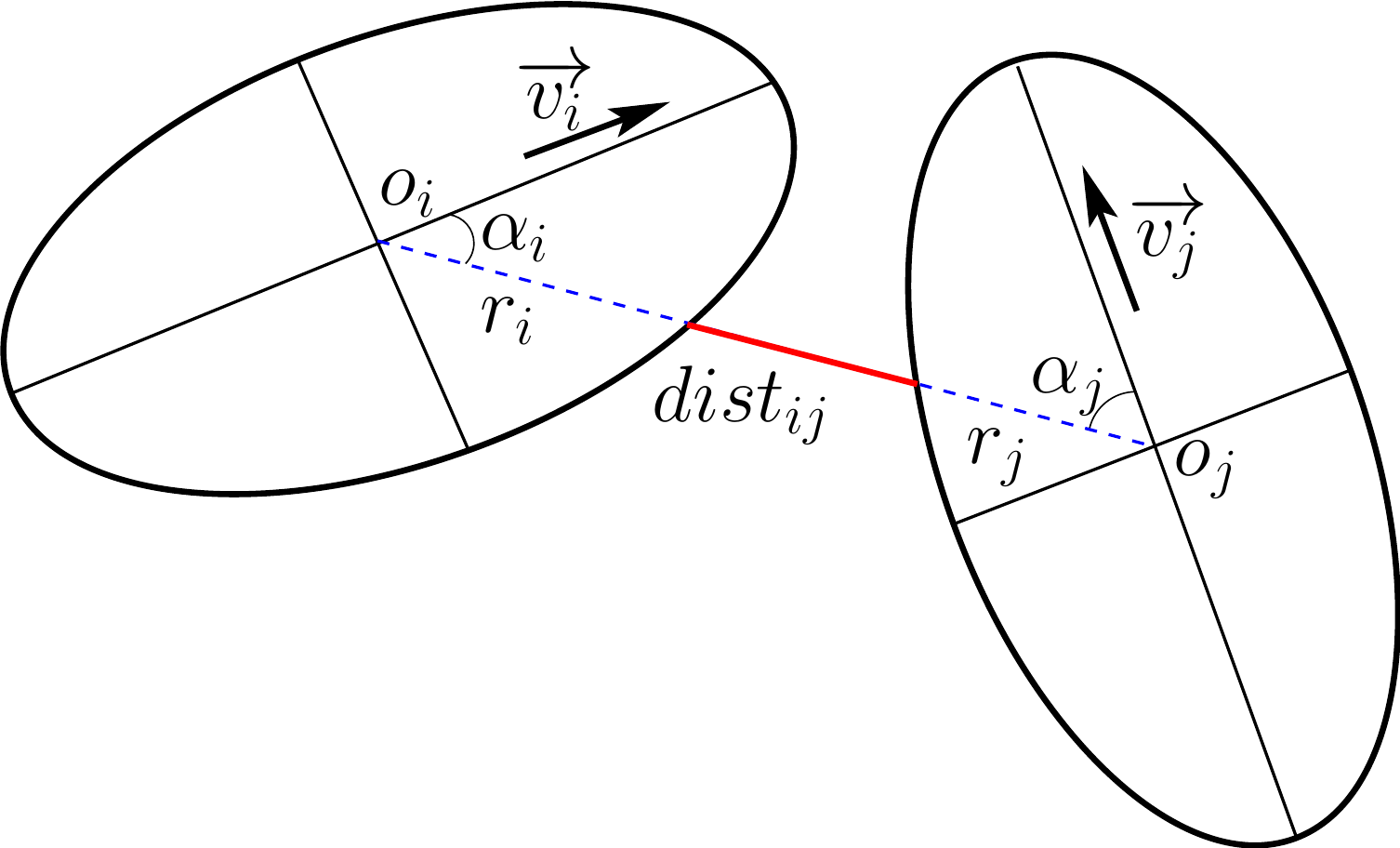}
\caption{(Color online) $\operatorname{dist}_{ij}$ is the distance between the borders of the  ellipses $i$ and $j$ along a line connecting their centers.}
\label{fig:dist}
\end{center}
\end{figure}

In this appendix we give details about the calculation of the distance
$\operatorname{dist}_{ij}$ between two ellipses which is defined as the distance 
between the borders of the ellipses, along a line connecting their centers (Fig.~\ref{fig:dist}).

By proper choice of the coordinate system the ellipse $i$ may be written
as quadratic form,

\begin{equation}
   \frac{x^2}{a_i^2} +  \frac{y^2}{b_i^2} = 1\,.
\label{eq:ellipse}
\end{equation}
In polar coordinates, with the origin at the center of the ellipse and
with the angular coordinate $\alpha_i$ 
measured from the major axis, one gets
\begin{equation}
  x = r_i  \cos(\alpha_i)\,,\qquad  y = r_i \sin(\alpha_i)\,.
\label{eq:alpha}
\end{equation}
By replacing the expressions of $x$\ and $y$ in Eq.~(\ref{eq:ellipse})
and rearranging we obtain the expression 
\begin{equation}
qr^2_i - 1 =0,
\label{eq:quad}
\end{equation}
for the polar radius $r_i$ with
\begin{equation}
q=\frac{\cos^2\alpha_i}{a_i^2} + \frac{\sin^2\alpha_i}{b_i^2}.
\end{equation}

In the same manner, we determine the polar radius $r_j$.

Finally, the distance $\operatorname{dist}_{ij}$ between the centers
of the ellipses $i$\ and $j$ is determined as follows (Fig.~\ref{fig:dist}):
\begin{equation}
  \operatorname{dist}_{ij}=\parallel \overrightarrow{o_io_j} \parallel
  -r_i -r_j\,.
\label{eq:dist}
\end{equation}

%%%%%%%%%%%%%%%%%%%%%%%%%%%%%%%%%%%%%%%
%%%%%%%%%%%%%%%%%%%%%%%%%%%%%%%%%%%%%%%

Note that the distance between two ellipses can be non-zero even when the ellipses touch or overlap.

%%%%%%%%%%%%%%%%%%%%%%%%%%%%%%%%%%%%%%%%%%%%%%%%%%%%%%%%%%%%%%%%%%%%%%%%%%

\section {Distance of closest approach}
\label{sec:dist-wall}
\begin{figure}[ht]
\begin{center}
 \subfigure{\label{fig:mindes}}
\includegraphics[width=0.75\columnwidth]{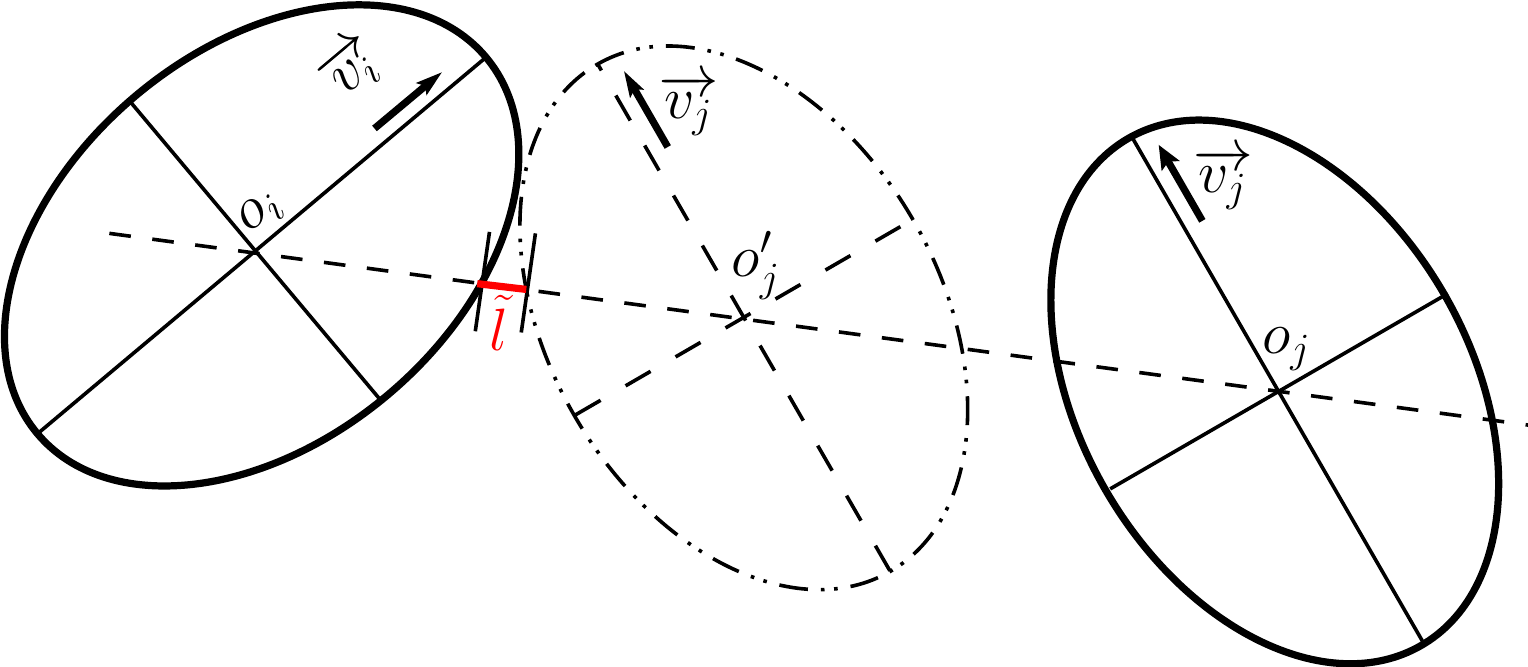}
\vspace{0.5cm}
\subfigure{\label{fig:mindel}}
\includegraphics[width=0.75\columnwidth]{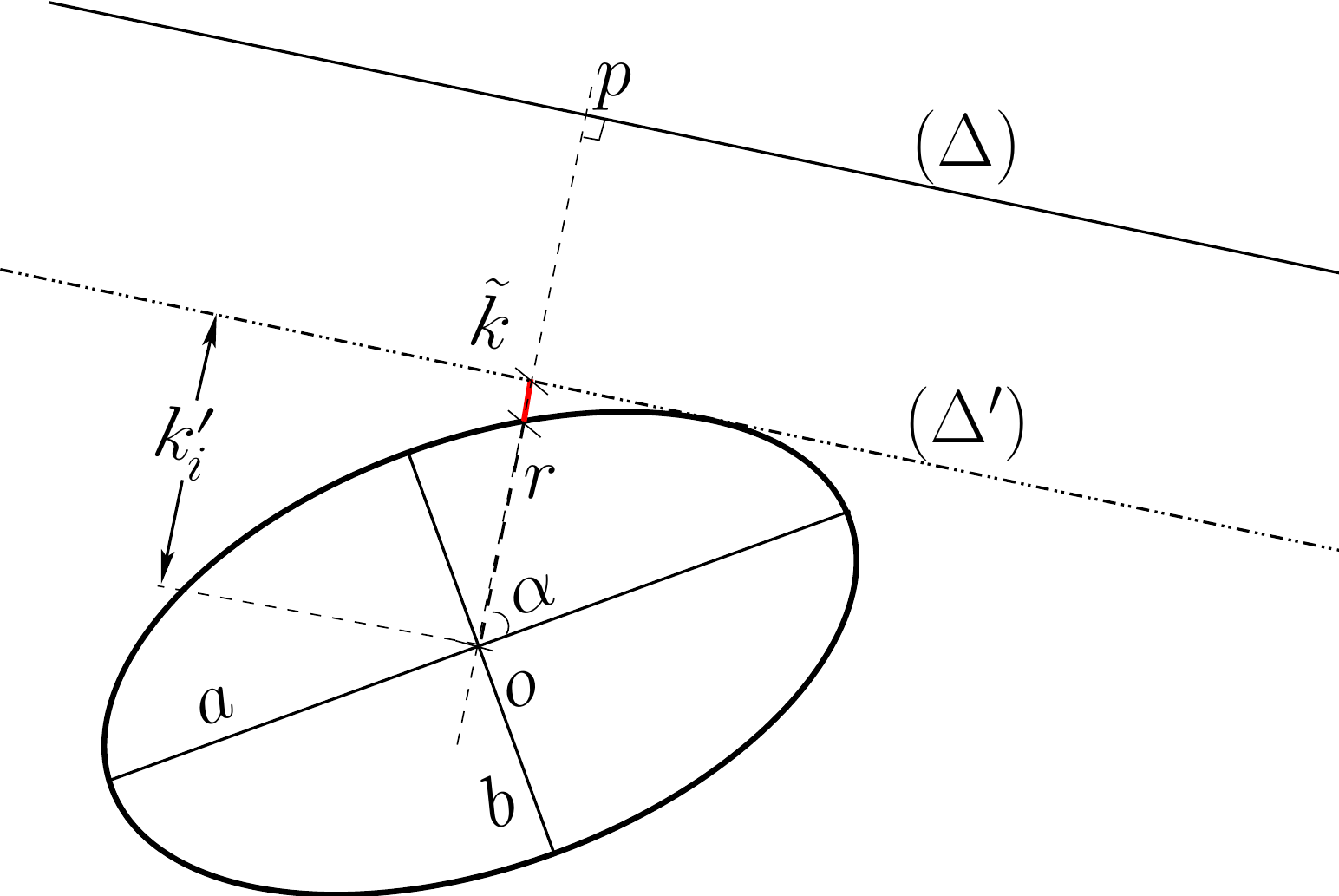}
\caption{(Color online) Top: Distance of closest approach of two ellipses. Bottom: Distance of closest approach between an ellipse and a line.
}
\label{fig:mindel-mindes}
\end{center}
\end{figure}

Distance of closest approach of two ellipses is the smallest distance between their borders, along a line connecting their centers while they are not overlapping. See Fig.~(\ref{fig:mindel-mindes}) top. To mitigate overlapping 
the repulsive forces are high for distances in a certain neighborhood 
of the distance of closest approach, see $\tilde{l}$ in Fig. \ref{fig:interp-dir}. An analytical solution of this distance for two arbitrary ellipses is presented in \cite{Zheng2007}.

In this appendix we describe an algorithm to calculate the distance of closest approach of an ellipse and a line  $(\Delta)$, which is the distance between the border of the ellipse, along a line connecting its center $o$ and the closest point on the line to $o$. For this purpose consider
without loss of generality an ellipse $i$ in canonical position and
let $(\Delta^\prime)$ be the line tangential to the ellipse $i$ and
parallel to $(\Delta)$ (Fig.~(\ref{fig:mindel-mindes}) bottom:
\begin{equation}
  (\Delta): y=cx+d\,,\;\;\;\;\;\; 
  (\Delta^\prime): y=cx+d^\prime\,.
\label{eq:walls}
\end{equation}
with known coefficients $c$ and $d$. 

To determine $d^\prime$ we solve the intersection equations of an
ellipse and a line, which yields the quadratic equation
\begin{equation}
q^\prime x^2 + p^\prime x + s^\prime =0,
\label{eq:quad2}
\end{equation}
with
\begin{equation}
q^\prime = \frac{1}{a^2}+ \frac{c^2}{b^2},\qquad
p^\prime= \frac{2cd^\prime}{b^2}\qquad \text{and}\;\;\;
s^\prime= \frac{{d^\prime}^2}{b^2}-1.
\end{equation}
As $(\Delta^\prime)$ is tangential to the ellipse we have
\begin{equation}
D=0
\label{eq:discrimant}
\end{equation}
with $D$ the discriminant of Eq.~(\ref{eq:quad2}). Solving
(\ref{eq:discrimant}) gives
\begin{equation}
d^\prime=\pm\sqrt{b^2 + a^2c^2}.
\label{eq:d}
\end{equation}

Finally the distance of closest approach of the ellipse $i$ and line $(\Delta)$ is

\begin{equation}
  \tilde{k} = k_i^\prime - r_i,
\label{eq:distW1}
\end{equation}
with $k_i^\prime$ the distance of $c_i$ to $(\Delta^\prime)$ and $r_i$
the polar radius as determined in Eq.~(\ref{eq:quad}).

%%%%%%%%%%%%%%%%%%%%%%%%%%%%%%%%%%%%%%%%%%%%%%%%%%%%%%%%%%%%%%%%%%%%%%%%%%

\section{Measurement method}
\label{sec:meas}

The mean velocity of pedestrian $i$ that enters the measurement are at
$(x^\text{in}_i,y^\text{in}_i)$ and leaves it at 
$(x^\text{out}_i,y^\text{out}_i)$ is determined as
\begin{equation}
  v_i = \frac{\sqrt{(x^\text{out}_i - x^\text{in}_i)^2 + (y^\text{out}_i -
      y^\text{in}_i)^2} }{t^\text{out}_i - t^\text{in}_i}
\label{eq:vel-2d}
\end{equation}
where $t^\text{in}_i$ is the entrance time and  $t^\text{out}_i$ exit 
time of $i$. For the one-dimensional case $y^\text{in}_i=y^\text{out}_i=0$.

The density is defined as follows: 
\begin{equation}
   \rho_i = \frac{1}{t^\text{out}_i -
     t^\text{in}_i}\int_{t_\text{in}}^{t_\text{out}}\!\rho (t)\, dt
\label{eq:rho-2d-i}
\end{equation}
\begin{equation}
   \rho (t) = \frac{N_\text{in}(t)}{l_m}.
\label{eq:rho-2d}
\end{equation}
with $l_m=2m$ the length of the measurement area in the movement
direction and $N_\text{in}(t)$ is the number of pedestrians within the area
at time $t$. In one dimensional space the measurement area is reduced to a measurement segment of length $l_m$.

%%%%%%%%%%%%%%%%%%%%%%%%%%%%%%%%%%%%%%%%%%%%%%%%%%%%%%%%%%%%%%%%%%%%%%%%%%

\end{document}